\documentclass[reprint,aps,prl]{revtex4-2}
\usepackage{amsmath,amssymb,amsfonts,dsfont}
\usepackage{graphicx}
\usepackage{color}
\usepackage{hyperref}
\usepackage{subfigure}
\usepackage{dcolumn}% Align table columns on decimal point
\usepackage{bm}% bold math
\usepackage{float}
\usepackage{mathtools}
\usepackage{amsthm}
\usepackage{bbm}
\usepackage{pxfonts}
\usepackage{pstricks}
\usepackage{verbatim}

\definecolor{orange}{cmyk}{0,0.5,1,0}
\definecolor{rossoCP3}{cmyk}{0,.88,.77,.40}
\definecolor{graa}{rgb}{0.8,0.8,0.8}
\definecolor{blaa}{rgb}{0.2,0.2,0.6}
\hypersetup{
	colorlinks, 
	bookmarksopen, 
	bookmarksnumbered,
	citecolor=blaa, 		%color of links to bibliography
	linkcolor=rossoCP3,	%color of internal links
	urlcolor=rossoCP3,			%color of external links 
	    }
\newcommand{\beq}{\begin{eqnarray}}
\newcommand{\eeq}{\end{eqnarray}}

\newcommand{\SU}{\mathrm{SU}}
\newcommand{\Sp}{\mathrm{Sp}}

\begin{document}
\title{Comment to ``The asymptotically-free gauge theories''}

\author{Giacomo Cacciapaglia}
\email{cacciapa@lpthe.jussieu.fr}
\affiliation{Laboratoire de Physique Theorique et Hautes Energies {\color{rossoCP3}LPTHE}, UMR 7589, Sorbonne Universit\'e \& CNRS, 4 place Jussieu, 75252 Paris Cedex 05, France.}
%\author{Konstantinos Kollias}
%\email{kkollias@clipper.ens.psl.eu}
%\affiliation{Laboratoire de Physique Theorique et Hautes Energies {\color{rossoCP3}LPTHE}, UMR 7589, Sorbonne Universit\'e \& CNRS, 4 place Jussieu, 75252 Paris Cedex 05, France.}
\author{Aldo Deandrea}
\email{deandrea@ip2i.in2p3.fr} 
\affiliation{Universit{\'e} Claude Bernard Lyon 1, {\color{rossoCP3}IP2I} UMR 5822, CNRS/IN2P3, 4 rue Enrico Fermi, 69622 Villeurbanne Cedex, France;\\
Department of Physics, University of Johannesburg,
PO Box 524, Auckland Park 2006, South Africa.}
\author{Francesco Sannino}
\email{sannino@qtc.sdu.dk}
\affiliation{{\color{rossoCP3}{$\hbar$}QTC} \& the Danish Institute for Advanced Study {\color{rossoCP3}\rm{Danish IAS}},  University of Southern Denmark, Campusvej 55, DK-5230 Odense M, Denmark;}
\affiliation{Dept. of Physics E. Pancini, Universit\`a di Napoli Federico II, via Cintia, 80126 Napoli, Italy;}
\affiliation{INFN sezione di Napoli, via Cintia, 80126 Napoli, Italy}
\affiliation{Scuola Superiore Meridionale, Largo S. Marcellino, 10, 80138 Napoli, Italy,}
%%%%%%%%%%%%%%%%%%%%%%%%%%%%%%%%%%%%%%%%%%%%%%%%%%%%%%%%%%%%%%%%%%%%%%%%%%

\begin{abstract}
The recent paper "\emph{The asymptotically-free gauge theories}" by Ben Gripaios and Khoi Le Nguyen Nguyen \cite{Gripaios:2025kcb} presents a proposed classification of gauge theories valid down to arbitrarily short scales. In this comment, we aim to clarify several points and address some statements that may be misleading. We also provide additional context by discussing relevant prior literature and existing classifications.
\end{abstract}

\maketitle

\section{Gauge anomaly cancellation}

The cancellation of gauge anomalies is essential in defining a consistent gauge theory. Hence, imposing the absence of ABJ anomalies \cite{Adler:1969gk,Bell:1969ts} and topological anomalies \cite{Witten:1982fp,Wang:2018qoy} is a central ingredient in any classification attempt. The former are relevant for $\SU(N)$ theories with complex representations, while the latter for $\Sp(2N)$ theories with pseudo-real irreducible representations (irreps). This analysis has been performed in \cite{Cacciapaglia:2025nff} to classify asymptotic-free (AF) gauge theories, showing that crucial restrictions are imposed on the available irreps. For instance, the three-index symmetric of $\SU(N)$, $(3,0,
\dots)$ in dynkin representation, cannot be used to define a  chiral AF theory \cite{Cacciapaglia:2025nff} for all $N$, while a one-family vector-like theory exists only for $N=2,3,4$. For the $(1,0,1,0,0)$ of $\SU(6)$, neither chiral nor vector-like theories can be defined. Furthermore, the well known Witten anomalies forbid the presence of an odd number of certain representations of $\Sp(2N)$ (a discussion of the Witten anomalies for general pseudo-real representations of $\Sp(2N)$ is present in the supplementary material of \cite{Cacciapaglia:2025nff}). 

The above constraints were not taken into account in \cite{Gripaios:2025kcb}, where anomalies are only briefly mentioned as reducing the number of viable theories for $\SU(5)$ (Lie algebra $A_4$). In fact, Ref.\cite{Cacciapaglia:2025nff} provides a `basis' of chiral families, which can be used to construct the whole class of AF anomaly-free $\SU(N)$ theories, when supplemented with vector-like fermions.

\section{Classifications present in the literature}

The statement ``\emph{here we provide a first classification of such theories, focussing on the case in which any matter is fermionic; the generalization to theories with scalar particles or supersymmetry will become obvious}'' is incorrect. For instance, vector-like AF theories based on $\SU(N)$ have been classified in \cite{Dietrich:2006cm}. The chiral case has been classified in \cite{Cacciapaglia:2025nff} by providing a basis of chiral families for all $\SU(N)$ and anomaly-free AF representations for $\Sp(2N)$. The classification of AF irreps of $\SU(N)$ has been pioneered by E.Eichten and F.Feinberg: Ref.~\cite{Eichten:1981mu} contains a more complete list than the one in Ref.~\cite{Eichten:1982pn} (the latter cited in \cite{Gripaios:2025kcb}).

Studying AF in the presence of scalars is not a mere generalization of the fermionic case (see next paragraph).

\section{Complete freedom with scalars}

Achieving AF in gauge theories with scalars is far from obvious. This is due to the presence of additional marginal operators, namely quartic couplings and Yukawas (with fermions). A simple analysis of the one-loop gauge beta function, as stated in \cite{Gripaios:2025kcb}, is far from enough: in fact, the sensitivity to the Yukawa couplings enters at two loops while that to the quartics at three loops. This order is dictated by the need to respect the Weyl consistency conditions \cite{Antipin:2013sga,Osborn:1989td,Jack:1990eb,Osborn:1991gm}. Complete asymptotic freedom, therefore, requires that all the couplings run free \cite{Callaway:1988ya,Holdom:2014hla,Giudice:2014tma,Pica:2016krb}.

\section{You can be safe rather than free}

The naive statement that ``\emph{asymptotically-free gauge theories provide the only known examples of interacting quantum field theories in four spacetime dimensions that are valid down to arbitrarily short scales}'' is challenged by the discovery of interactive four-dimensional gauge-Yukawa asymptotically safe theories \cite{Litim:2014uca}. 

   At its core, this discovery demonstrates the existence of interacting ultraviolet (UV) fixed points in non-supersymmetric gauge-Yukawa theories. In simpler terms, it showed that it is possible for the interaction strengths of a quantum field theory not to grow infinitely large at very high energies, a fatal flaw known as a Landau pole, but instead to approach a constant, non-zero value. This ``asymptotic safety'' ensures that the theory remains well behaved and predictive. Hence, the naive expectation, largely shaped by the success of Quantum Chromodynamics (QCD), that only "asymptotically free" theories can be considered fundamental, is incorrect since this discovery provides a compelling alternative. Interactions can persist at the highest energy scales without leading to inconsistencies.

\bibliography{biblio}

\begin{thebibliography}{18}%
\makeatletter
\providecommand \@ifxundefined [1]{%
 \@ifx{#1\undefined}
}%
\providecommand \@ifnum [1]{%
 \ifnum #1\expandafter \@firstoftwo
 \else \expandafter \@secondoftwo
 \fi
}%
\providecommand \@ifx [1]{%
 \ifx #1\expandafter \@firstoftwo
 \else \expandafter \@secondoftwo
 \fi
}%
\providecommand \natexlab [1]{#1}%
\providecommand \enquote  [1]{``#1''}%
\providecommand \bibnamefont  [1]{#1}%
\providecommand \bibfnamefont [1]{#1}%
\providecommand \citenamefont [1]{#1}%
\providecommand \href@noop [0]{\@secondoftwo}%
\providecommand \href [0]{\begingroup \@sanitize@url \@href}%
\providecommand \@href[1]{\@@startlink{#1}\@@href}%
\providecommand \@@href[1]{\endgroup#1\@@endlink}%
\providecommand \@sanitize@url [0]{\catcode `\\12\catcode `\$12\catcode `\&12\catcode `\#12\catcode `\^12\catcode `\_12\catcode `\%12\relax}%
\providecommand \@@startlink[1]{}%
\providecommand \@@endlink[0]{}%
\providecommand \url  [0]{\begingroup\@sanitize@url \@url }%
\providecommand \@url [1]{\endgroup\@href {#1}{\urlprefix }}%
\providecommand \urlprefix  [0]{URL }%
\providecommand \Eprint [0]{\href }%
\providecommand \doibase [0]{https://doi.org/}%
\providecommand \selectlanguage [0]{\@gobble}%
\providecommand \bibinfo  [0]{\@secondoftwo}%
\providecommand \bibfield  [0]{\@secondoftwo}%
\providecommand \translation [1]{[#1]}%
\providecommand \BibitemOpen [0]{}%
\providecommand \bibitemStop [0]{}%
\providecommand \bibitemNoStop [0]{.\EOS\space}%
\providecommand \EOS [0]{\spacefactor3000\relax}%
\providecommand \BibitemShut  [1]{\csname bibitem#1\endcsname}%
\let\auto@bib@innerbib\@empty
%</preamble>
\bibitem [{\citenamefont {Gripaios}\ and\ \citenamefont {Le~Nguyen~Nguyen}(2025)}]{Gripaios:2025kcb}%
  \BibitemOpen
  \bibfield  {author} {\bibinfo {author} {\bibfnamefont {B.}~\bibnamefont {Gripaios}}\ and\ \bibinfo {author} {\bibfnamefont {K.}~\bibnamefont {Le~Nguyen~Nguyen}},\ }\bibfield  {title} {\bibinfo {title} {{The asymptotically-free gauge theories}},\ }\href@noop {} {\  (\bibinfo {year} {2025})},\ \Eprint {https://arxiv.org/abs/2507.12348} {arXiv:2507.12348 [hep-th]} \BibitemShut {NoStop}%
\bibitem [{\citenamefont {Adler}(1969)}]{Adler:1969gk}%
  \BibitemOpen
  \bibfield  {author} {\bibinfo {author} {\bibfnamefont {S.~L.}\ \bibnamefont {Adler}},\ }\bibfield  {title} {\bibinfo {title} {{Axial vector vertex in spinor electrodynamics}},\ }\href {https://doi.org/10.1103/PhysRev.177.2426} {\bibfield  {journal} {\bibinfo  {journal} {Phys. Rev.}\ }\textbf {\bibinfo {volume} {177}},\ \bibinfo {pages} {2426} (\bibinfo {year} {1969})}\BibitemShut {NoStop}%
\bibitem [{\citenamefont {Bell}\ and\ \citenamefont {Jackiw}(1969)}]{Bell:1969ts}%
  \BibitemOpen
  \bibfield  {author} {\bibinfo {author} {\bibfnamefont {J.~S.}\ \bibnamefont {Bell}}\ and\ \bibinfo {author} {\bibfnamefont {R.}~\bibnamefont {Jackiw}},\ }\bibfield  {title} {\bibinfo {title} {{A PCAC puzzle: $\pi^0 \to \gamma \gamma$ in the $\sigma$ model}},\ }\href {https://doi.org/10.1007/BF02823296} {\bibfield  {journal} {\bibinfo  {journal} {Nuovo Cim. A}\ }\textbf {\bibinfo {volume} {60}},\ \bibinfo {pages} {47} (\bibinfo {year} {1969})}\BibitemShut {NoStop}%
\bibitem [{\citenamefont {Witten}(1982)}]{Witten:1982fp}%
  \BibitemOpen
  \bibfield  {author} {\bibinfo {author} {\bibfnamefont {E.}~\bibnamefont {Witten}},\ }\bibfield  {title} {\bibinfo {title} {{An SU(2) Anomaly}},\ }\href {https://doi.org/10.1016/0370-2693(82)90728-6} {\bibfield  {journal} {\bibinfo  {journal} {Phys. Lett. B}\ }\textbf {\bibinfo {volume} {117}},\ \bibinfo {pages} {324} (\bibinfo {year} {1982})}\BibitemShut {NoStop}%
\bibitem [{\citenamefont {Wang}\ \emph {et~al.}(2019)\citenamefont {Wang}, \citenamefont {Wen},\ and\ \citenamefont {Witten}}]{Wang:2018qoy}%
  \BibitemOpen
  \bibfield  {author} {\bibinfo {author} {\bibfnamefont {J.}~\bibnamefont {Wang}}, \bibinfo {author} {\bibfnamefont {X.-G.}\ \bibnamefont {Wen}},\ and\ \bibinfo {author} {\bibfnamefont {E.}~\bibnamefont {Witten}},\ }\bibfield  {title} {\bibinfo {title} {{A New SU(2) Anomaly}},\ }\href {https://doi.org/10.1063/1.5082852} {\bibfield  {journal} {\bibinfo  {journal} {J. Math. Phys.}\ }\textbf {\bibinfo {volume} {60}},\ \bibinfo {pages} {052301} (\bibinfo {year} {2019})},\ \Eprint {https://arxiv.org/abs/1810.00844} {arXiv:1810.00844 [hep-th]} \BibitemShut {NoStop}%
\bibitem [{\citenamefont {Cacciapaglia}\ \emph {et~al.}(2025)\citenamefont {Cacciapaglia}, \citenamefont {Deandrea}, \citenamefont {Kollias},\ and\ \citenamefont {Sannino}}]{Cacciapaglia:2025nff}%
  \BibitemOpen
  \bibfield  {author} {\bibinfo {author} {\bibfnamefont {G.}~\bibnamefont {Cacciapaglia}}, \bibinfo {author} {\bibfnamefont {A.}~\bibnamefont {Deandrea}}, \bibinfo {author} {\bibfnamefont {K.}~\bibnamefont {Kollias}},\ and\ \bibinfo {author} {\bibfnamefont {F.}~\bibnamefont {Sannino}},\ }\bibfield  {title} {\bibinfo {title} {{Grand-unification Theory Atlas: Standard Model and Beyond}},\ }\href@noop {} {\  (\bibinfo {year} {2025})},\ \Eprint {https://arxiv.org/abs/2507.06368} {arXiv:2507.06368 [hep-ph]} \BibitemShut {NoStop}%
\bibitem [{\citenamefont {Dietrich}\ and\ \citenamefont {Sannino}(2007)}]{Dietrich:2006cm}%
  \BibitemOpen
  \bibfield  {author} {\bibinfo {author} {\bibfnamefont {D.~D.}\ \bibnamefont {Dietrich}}\ and\ \bibinfo {author} {\bibfnamefont {F.}~\bibnamefont {Sannino}},\ }\bibfield  {title} {\bibinfo {title} {{Conformal window of SU(N) gauge theories with fermions in higher dimensional representations}},\ }\href {https://doi.org/10.1103/PhysRevD.75.085018} {\bibfield  {journal} {\bibinfo  {journal} {Phys. Rev. D}\ }\textbf {\bibinfo {volume} {75}},\ \bibinfo {pages} {085018} (\bibinfo {year} {2007})},\ \Eprint {https://arxiv.org/abs/hep-ph/0611341} {arXiv:hep-ph/0611341} \BibitemShut {NoStop}%
\bibitem [{\citenamefont {Eichten}\ and\ \citenamefont {Feinberg}(1982)}]{Eichten:1981mu}%
  \BibitemOpen
  \bibfield  {author} {\bibinfo {author} {\bibfnamefont {E.}~\bibnamefont {Eichten}}\ and\ \bibinfo {author} {\bibfnamefont {F.}~\bibnamefont {Feinberg}},\ }\bibfield  {title} {\bibinfo {title} {{Comment on Tumbling Gauge Theories}},\ }\href {https://doi.org/10.1016/0370-2693(82)91242-4} {\bibfield  {journal} {\bibinfo  {journal} {Phys. Lett. B}\ }\textbf {\bibinfo {volume} {110}},\ \bibinfo {pages} {232} (\bibinfo {year} {1982})}\BibitemShut {NoStop}%
\bibitem [{\citenamefont {Eichten}\ \emph {et~al.}(1982)\citenamefont {Eichten}, \citenamefont {Kang},\ and\ \citenamefont {Koh}}]{Eichten:1982pn}%
  \BibitemOpen
  \bibfield  {author} {\bibinfo {author} {\bibfnamefont {E.}~\bibnamefont {Eichten}}, \bibinfo {author} {\bibfnamefont {K.}~\bibnamefont {Kang}},\ and\ \bibinfo {author} {\bibfnamefont {I.-G.}\ \bibnamefont {Koh}},\ }\bibfield  {title} {\bibinfo {title} {{Anomaly Free Complex Representations in SU(N)}},\ }\href {https://doi.org/10.1063/1.525299} {\bibfield  {journal} {\bibinfo  {journal} {J. Math. Phys.}\ }\textbf {\bibinfo {volume} {23}},\ \bibinfo {pages} {2529} (\bibinfo {year} {1982})}\BibitemShut {NoStop}%
\bibitem [{\citenamefont {Antipin}\ \emph {et~al.}(2013)\citenamefont {Antipin}, \citenamefont {Gillioz}, \citenamefont {Krog}, \citenamefont {M{\o}lgaard},\ and\ \citenamefont {Sannino}}]{Antipin:2013sga}%
  \BibitemOpen
  \bibfield  {author} {\bibinfo {author} {\bibfnamefont {O.}~\bibnamefont {Antipin}}, \bibinfo {author} {\bibfnamefont {M.}~\bibnamefont {Gillioz}}, \bibinfo {author} {\bibfnamefont {J.}~\bibnamefont {Krog}}, \bibinfo {author} {\bibfnamefont {E.}~\bibnamefont {M{\o}lgaard}},\ and\ \bibinfo {author} {\bibfnamefont {F.}~\bibnamefont {Sannino}},\ }\bibfield  {title} {\bibinfo {title} {{Standard Model Vacuum Stability and Weyl Consistency Conditions}},\ }\href {https://doi.org/10.1007/JHEP08(2013)034} {\bibfield  {journal} {\bibinfo  {journal} {JHEP}\ }\textbf {\bibinfo {volume} {08}},\ \bibinfo {pages} {034}},\ \Eprint {https://arxiv.org/abs/1306.3234} {arXiv:1306.3234 [hep-ph]} \BibitemShut {NoStop}%
\bibitem [{\citenamefont {Osborn}(1989)}]{Osborn:1989td}%
  \BibitemOpen
  \bibfield  {author} {\bibinfo {author} {\bibfnamefont {H.}~\bibnamefont {Osborn}},\ }\bibfield  {title} {\bibinfo {title} {{Derivation of a Four-dimensional $c$ Theorem}},\ }\href {https://doi.org/10.1016/0370-2693(89)90729-6} {\bibfield  {journal} {\bibinfo  {journal} {Phys. Lett. B}\ }\textbf {\bibinfo {volume} {222}},\ \bibinfo {pages} {97} (\bibinfo {year} {1989})}\BibitemShut {NoStop}%
\bibitem [{\citenamefont {Jack}\ and\ \citenamefont {Osborn}(1990)}]{Jack:1990eb}%
  \BibitemOpen
  \bibfield  {author} {\bibinfo {author} {\bibfnamefont {I.}~\bibnamefont {Jack}}\ and\ \bibinfo {author} {\bibfnamefont {H.}~\bibnamefont {Osborn}},\ }\bibfield  {title} {\bibinfo {title} {{Analogs for the $c$ Theorem for Four-dimensional Renormalizable Field Theories}},\ }\href {https://doi.org/10.1016/0550-3213(90)90584-Z} {\bibfield  {journal} {\bibinfo  {journal} {Nucl. Phys. B}\ }\textbf {\bibinfo {volume} {343}},\ \bibinfo {pages} {647} (\bibinfo {year} {1990})}\BibitemShut {NoStop}%
\bibitem [{\citenamefont {Osborn}(1991)}]{Osborn:1991gm}%
  \BibitemOpen
  \bibfield  {author} {\bibinfo {author} {\bibfnamefont {H.}~\bibnamefont {Osborn}},\ }\bibfield  {title} {\bibinfo {title} {{Weyl consistency conditions and a local renormalization group equation for general renormalizable field theories}},\ }\href {https://doi.org/10.1016/0550-3213(91)80030-P} {\bibfield  {journal} {\bibinfo  {journal} {Nucl. Phys. B}\ }\textbf {\bibinfo {volume} {363}},\ \bibinfo {pages} {486} (\bibinfo {year} {1991})}\BibitemShut {NoStop}%
\bibitem [{\citenamefont {Callaway}(1988)}]{Callaway:1988ya}%
  \BibitemOpen
  \bibfield  {author} {\bibinfo {author} {\bibfnamefont {D.~J.~E.}\ \bibnamefont {Callaway}},\ }\bibfield  {title} {\bibinfo {title} {{Triviality Pursuit: Can Elementary Scalar Particles Exist?}},\ }\href {https://doi.org/10.1016/0370-1573(88)90008-7} {\bibfield  {journal} {\bibinfo  {journal} {Phys. Rept.}\ }\textbf {\bibinfo {volume} {167}},\ \bibinfo {pages} {241} (\bibinfo {year} {1988})}\BibitemShut {NoStop}%
\bibitem [{\citenamefont {Holdom}\ \emph {et~al.}(2015)\citenamefont {Holdom}, \citenamefont {Ren},\ and\ \citenamefont {Zhang}}]{Holdom:2014hla}%
  \BibitemOpen
  \bibfield  {author} {\bibinfo {author} {\bibfnamefont {B.}~\bibnamefont {Holdom}}, \bibinfo {author} {\bibfnamefont {J.}~\bibnamefont {Ren}},\ and\ \bibinfo {author} {\bibfnamefont {C.}~\bibnamefont {Zhang}},\ }\bibfield  {title} {\bibinfo {title} {{Stable Asymptotically Free Extensions (SAFEs) of the Standard Model}},\ }\href {https://doi.org/10.1007/JHEP03(2015)028} {\bibfield  {journal} {\bibinfo  {journal} {JHEP}\ }\textbf {\bibinfo {volume} {03}},\ \bibinfo {pages} {028}},\ \Eprint {https://arxiv.org/abs/1412.5540} {arXiv:1412.5540 [hep-ph]} \BibitemShut {NoStop}%
\bibitem [{\citenamefont {Giudice}\ \emph {et~al.}(2015)\citenamefont {Giudice}, \citenamefont {Isidori}, \citenamefont {Salvio},\ and\ \citenamefont {Strumia}}]{Giudice:2014tma}%
  \BibitemOpen
  \bibfield  {author} {\bibinfo {author} {\bibfnamefont {G.~F.}\ \bibnamefont {Giudice}}, \bibinfo {author} {\bibfnamefont {G.}~\bibnamefont {Isidori}}, \bibinfo {author} {\bibfnamefont {A.}~\bibnamefont {Salvio}},\ and\ \bibinfo {author} {\bibfnamefont {A.}~\bibnamefont {Strumia}},\ }\bibfield  {title} {\bibinfo {title} {{Softened Gravity and the Extension of the Standard Model up to Infinite Energy}},\ }\href {https://doi.org/10.1007/JHEP02(2015)137} {\bibfield  {journal} {\bibinfo  {journal} {JHEP}\ }\textbf {\bibinfo {volume} {02}},\ \bibinfo {pages} {137}},\ \Eprint {https://arxiv.org/abs/1412.2769} {arXiv:1412.2769 [hep-ph]} \BibitemShut {NoStop}%
\bibitem [{\citenamefont {Pica}\ \emph {et~al.}(2017)\citenamefont {Pica}, \citenamefont {Ryttov},\ and\ \citenamefont {Sannino}}]{Pica:2016krb}%
  \BibitemOpen
  \bibfield  {author} {\bibinfo {author} {\bibfnamefont {C.}~\bibnamefont {Pica}}, \bibinfo {author} {\bibfnamefont {T.~A.}\ \bibnamefont {Ryttov}},\ and\ \bibinfo {author} {\bibfnamefont {F.}~\bibnamefont {Sannino}},\ }\bibfield  {title} {\bibinfo {title} {{Conformal Phase Diagram of Complete Asymptotically Free Theories}},\ }\href {https://doi.org/10.1103/PhysRevD.96.074015} {\bibfield  {journal} {\bibinfo  {journal} {Phys. Rev. D}\ }\textbf {\bibinfo {volume} {96}},\ \bibinfo {pages} {074015} (\bibinfo {year} {2017})},\ \Eprint {https://arxiv.org/abs/1605.04712} {arXiv:1605.04712 [hep-th]} \BibitemShut {NoStop}%
\bibitem [{\citenamefont {Litim}\ and\ \citenamefont {Sannino}(2014)}]{Litim:2014uca}%
  \BibitemOpen
  \bibfield  {author} {\bibinfo {author} {\bibfnamefont {D.~F.}\ \bibnamefont {Litim}}\ and\ \bibinfo {author} {\bibfnamefont {F.}~\bibnamefont {Sannino}},\ }\bibfield  {title} {\bibinfo {title} {{Asymptotic safety guaranteed}},\ }\href {https://doi.org/10.1007/JHEP12(2014)178} {\bibfield  {journal} {\bibinfo  {journal} {JHEP}\ }\textbf {\bibinfo {volume} {12}},\ \bibinfo {pages} {178}},\ \Eprint {https://arxiv.org/abs/1406.2337} {arXiv:1406.2337 [hep-th]} \BibitemShut {NoStop}%
\end{thebibliography}%

\end{document}